\documentstyle[sprocl,epsf]{article}

\newcommand{\mrm}[1]{\mbox{\rm #1}}

\newcommand{\rfn}[1]{(\ref{#1})}
\newcommand{\db}{\hspace{-0.2ex}\not\hspace{-0.7ex}D\hspace{0.1ex}}

\newcommand{\beq}{\begin{equation}}
\newcommand{\eeq}{\end{equation}}
\newcommand{\bea}{\begin{eqnarray}}
\newcommand{\eea}{\end{eqnarray}}
\newcommand{\bes}{\begin{eqnarray*}}
\newcommand{\ees}{\end{eqnarray*}}   
\newcommand{\eq}[1]{eq.~(\ref{#1})}

\newcommand{\Eq}[1]{Eq.~(\ref{#1})}
\newcommand{\ea}{{\it et al.}}

\newcommand{\np}[1]{Nucl. Phys. {\bf #1}}
\newcommand{\pl}[1]{Phys. Lett. {\bf #1}}
\newcommand{\pr}[1]{Phys. Rev. {\bf #1}}
\newcommand{\prl}[1]{Phys. Rev. Lett. {\bf #1}}

\newcommand{\prep}[1]{Phys. Rep. {\bf #1}}

\def\lsim{\mathrel{\vcenter{\hbox{$<$}\nointerlineskip\hbox{$\sim$}}}}

\def\meg{$\mu \rightarrow e\ \gamma$}
\def\mec{$\mu$--$e$ conversion}

\begin{document}

\title{$\mu$--$e$ CONVERSION VERSUS $\mu \rightarrow e\ \gamma$}

\author{Martti Raidal} 

\address{Departament de F\'{\i}sica Te\`orica, IFIC, 
CSIC-Universitat de Val\`encia,\\ E-46100 Burjassot, Val\'encia, Spain}

\maketitle
\abstracts{
We compare two most relevant loop induced muon number violating
processes, $\mu \rightarrow e\ \gamma$  and
$\mu$--$e$ conversion in nuclei, in constraining new physics. 
Because of much richer structure of the latter process it may
be enhanced with respect to $\mu \rightarrow e\ \gamma$ 
by a large  $\ln(m^2_\mu/\Lambda^2),$ where $\Lambda$ is the scale 
of the new physics. After model-independent considerations we 
present two examples constraining the $R$-parity violating couplings 
of MSSM and the off-diagonal couplings of the doubly charged Higgs bosons
from the non-observation of \mec.}

\section{Introduction}

The precision reached in the last years in the experiments searching for
\mec{}  in nuclei at PSI \cite{psiti,psipb} 
and the expected improvements in the sensitivity 
of the experiments at PSI in the next years 
by more than two  orders of magnitude \cite{imprti} will make 
\mec{} the main test of muon flavour  conservation for
most of the extensions of the standard model (SM). 
Moreover, according to the recent BNL proposal \cite{proposal}
further improvements in the experimental sensitivity down to the level
$10^{-16}$ are feasible. The experimental prospects in this field 
have been recently reviewed by A. Czarnecki \cite{czarnecki}.

In any renormalizable model \meg\ can occur only at the loop level. 
For a wide class of models, discussed in the present talk, 
 this is also true for \mec\ (which, in principle,
can be also a tree level process).  If this is the
case, and given the present experimental accuracy, one often finds 
\cite{vergados} that 
the bounds on new physics coming from \meg{} 
are stronger than the bounds found from \mec{}.
However, this does not need to be always the case.
Indeed, it has been noticed already in the early works 
of Ref.~\cite{altarelli} 
that in some cases \mec{} constrains new physics more stringently than \meg{}. 
More recently it has been shown rigourosly \cite{RS} using the effective 
quantum field theory  in which class of models \mec{} 
is enhanced by large logarithms compared with \meg. 
Using this enhancement new strong bounds
have been derived for the doubly charged Higgs interactions 
\cite{RS} and for the $R$-parity violating couplings of the minimal 
supersymmetric standard model (MSSM) \cite{HMRS}.

\section{Effective Lagrangian description of \mec}

Assuming that the relevant physics responsible for muon-number
nonconservation occurs at some scale
$\Lambda > \Lambda_{F} \equiv$~Fermi~scale, we can write the relevant
effective Lagrangian at low energies as
\def\lla{{\cal L}^L}
\def\lra{{\cal L}^R}
\def\lsigmal{{\cal L}^{\sigma L}}
\def\lsigmar{{\cal L}^{\sigma R}}
\def\lsigma{{\cal L}^{\sigma}}
\def\lll{{\cal L}^{LL}}
\def\lrr{{\cal L}^{RR}}
\begin{equation}
\label{eq:leff}
{\cal L}_{eff} = \lla + \lsigmal + \lll + (L\to R)+ \cdots, 
\end{equation}
where
$\lla = \alpha^{L}_{ij}/((4\pi)^2 \Lambda^2) e\ \overline{e_{iL}} 
\gamma_\nu e_{jL} \partial_\mu F^{\mu\nu},$
$\lsigmal=
\alpha^{\sigma L}_{ij}/((4\pi)^2 \Lambda^2) e\ \overline{e_{iL}} 
\sigma_{\mu\nu}$ $ i \db e_{jL} F^{\mu\nu} + \mathrm{h.c.},$
$\lll=
\alpha^{LL}_{ik;lj}/\Lambda^2 
(\overline{e_{iL}} e_{kL}^c)(\overline{e_{lL}^c} e_{jL}) ,$
and similarly for the right-handed interactions.
We expect that the terms $\lla$, $\lra$, $\lsigmal$, $\lsigmar$ are generated
at one loop in the renormalizable theories since they cannot be obtained from
renormalizable vertices at tree level.
It is important to note that the Lagrangian \rfn{eq:leff} has to
be interpreted as a Lagrangian in the effective field theory 
approach. This means that four-fermion interactions, which
are generated at tree level, can be used at one loop and will 
generate non-analytical contributions to the electromagnetic
form factors.

Theory of \mec{} in nuclei was first studied by Weinberg and Feinberg 
in Ref. \cite{wf}. Since then
various nuclear models and approximations are used in the literature to 
calculate coherent \mec{} nuclear form factors.
We follow the notation of Ref.~\cite{chiang}.
The coherent \mec{} branching ratio $R_{\mu e}$ can be expressed as 
\cite{chiang}
\beq
R_{\mu e}=C\,\frac{8\alpha^5\,m_\mu^5\,Z^4_{eff}\,Z\,|\overline{F_p}(p_e)|^2}
{\Gamma_{capt}}
\cdot \frac{\xi_0^2}{q^4}\, ,
\label{mecrate}
\eeq
where $\xi_0^2=|f_{E0}+f_{M1}|^2+|f_{E1}+f_{M0}|^2$ contains the 
electromagnetic form factors and
$C^{Ti}=1.0,$ $C^{Pb}=1.4,$ $Z^{Ti}_{eff}=17.61,$ $Z^{Pb}_{eff}=33.81$,
$\Gamma_{capt}^{Ti}=2.59\cdot 10^6$~s$^{-1}$,
$\Gamma_{capt}^{Pb}=1.3\cdot 10^7$~s$^{-1}$ and the proton nuclear 
form factors are $\overline{F_p}^{Ti}(q)=0.55$ and
$\overline{F_p}^{Pb}(q)=0.25.$
One should note that the \meg{} branching ratio,
$R_\gamma=96\pi^3 \alpha/(G_F^2m_\mu^4)( |f_{M1}|^2+|f_{E1}|^2 ),$
depends on a different combination of the form factors.

We have computed the form factors at one loop level 
starting from the Lagrangian \rfn{eq:leff} in Ref. \cite{RS}.
There are new loop
contributions only to the form factors $f_{E0}$ and $f_{M0}$
but not to  $f_{E1}$ and $f_{M1}.$
Those contributions always contain a term which is proportional to 
$\ln(q^2/\Lambda^2)$ or $\ln(m_\tau^2/\Lambda^2).$
This term which is
completely {\it independent} of the details of the model that 
originate the four-fermion interaction gives a large 
enhancement for the form factors $f_{E0}$ and $f_{M0}$ while the 
enhancement is absent in the form factors $f_{E1}$ and $f_{M1}$.
To compare the branching ratios of \mec{} and \meg{}
we consider for definiteness only the right-handed operators in 
Lagrangian \rfn{eq:leff}.
Substituting  the form factors to \eq{mecrate}
(with  $\Lambda=1$ TeV in the logarithms)  we find  
\bea
R_{\mu e}& =& 1.2 \,(3.5) \cdot 10^{-5}\;\mrm{TeV}^{4}\;
\left(\frac{\alpha_{ke;k\mu}^{RR}}{\Lambda^2}\right)^2
\,,
\label{tirate}
\eea
where the first number corresponds to \mec\ in $Ti$ and the second
in $Pb$ and $Au.$
In the effective Lagrangian framework \meg{} does not get contributions 
from loops and assuming naturally  
$\alpha^{\sigma R}\equiv\alpha^{RR}$  we obtain
$R_{\gamma} = 1.2\cdot 10^{-5} \mrm{TeV}^{4}
(\alpha_{ke;k\mu}^{RR}/\Lambda^2)^2.$
Thus the rates of \mec\ and \meg\ in our models are comparable in magnitude.

\begin{table}
\begin{tabular}{|c|c|c|c|c|}
\hline\hline
$R=$ & $4.6\cdot 10^{-11}$ &$4.3\cdot 10^{-12}$&$5.0\cdot 10^{-13}$&
$3.0\cdot 10^{-14}$ \\
\hline\hline
log-enhanced $\mu$--$e$ & 
32 & 44 &101 & 158
\\
\hline
non-enhanced $\mu$--$e$ &
7 & 9 & 20 & 32 \\
\hline
$\mu\rightarrow e\gamma$ & 23 & 41 & 70 & 141
\\
\hline\hline
\end{tabular}
\caption{Values of  $\Lambda$, in TeV, probed in 
\mec{} and \meg{} for different
upper bounds on the branching ratios. The upper bound $4.6\cdot 10^{-11}$
is the present bound for \mec{} on $Pb$ and it is very close to the present
\meg{} bound ($4.9\cdot 10^{-11}$). $4.3\cdot 10^{-12}$ is the present
bound for \mec{} on $Ti$. $5\cdot 10^{-13}$ and $3\cdot 10^{-14}$ are
the expected bounds in the next year for \mec{} on $Au$ and $Ti$,
respectively.
}
\end{table}

The present experimental upper 
limits on the branching ratios of the processes are\cite{psiti,psipb,pdb}
$R^{Ti}_{\mu e}(exp)\lsim 4.3\cdot 10^{-12},$
$R^{Pb}_{\mu e}(exp)\lsim 4.6\cdot 10^{-11} $  and
$R_{\gamma}(exp)\lsim 4.9\cdot 10^{-11}.$ 
SINDRUM II experiment at PSI taking presently data  on gold will reach the 
sensitivity\cite{imprti} $R^{Au,\, expected}_{\mu e}\lsim 5\cdot 10^{-13}$
and starting next year the final run on $Ti$ it should reach\cite{imprti}
 $R^{Ti,\, expected}_{\mu e}\lsim 3\cdot 10^{-14}.$ 
To show which scales of new physics $\Lambda$ can be probed in
\mec{} and \meg{} experiments we have presented the values of $\Lambda$
in TeV-s in Table 1 for different experimental upper bounds
on the branching ratios  of the processes. 
All the couplings $\alpha$ are  taken to 
be equal to unity. We have considered both classes of models
with and without logarithmic enhancement of \mec{}.   
If the experimental limits for \mec{} and \meg{} are equal then 
\mec{} enhanced by large logarithms has 
better sensitivity to $\Lambda$ than \meg{}, especially 
in the case of $Pb$ and $Au$ experiments.

\section{Models with doubly charged Higgses}
\label{sec:model}

As the first example 
let us consider an extension of the
SM by adding just a doubly charged  scalar singlet $\kappa^{++}$ \cite{babu}.
However, the limits we derive apply with a good accuracy also for
the interactions of triplet scalars appearing in the models with enlarged
Higgs sectors  as well as in the left-right symmetric  models \cite{lr}.
This is because the doubly charged component
of the triplet gives the  dominant contribution both to \mec{} and \meg{}.

$\kappa^{++}$ coupling to right-handed leptons is described by  
\begin{equation}
{\cal L}_\kappa = h_{ij} \overline{e_{iR}^c} e_{jR}\,\kappa^{++} + 
\mathrm{h.c.}
\label{eq:lk}
\end{equation}
From this interaction we obtain easily
the four-fermion interaction \linebreak 
$1/m_\kappa^2 h^*_{ki} h_{lj} 
(\overline{e_{iR}} e_{kR}^c)(\overline{e_{lR}^c} e_{jR}).$
This interaction is of the type $\lrr$ and one can immediately identify
$\alpha_{ik;lj}^{RR} = h^*_{ik} h_{lj}$ and 
$\Lambda = m_\kappa.$
By using the $\overline{\mathrm{MS}}$ renormalization scheme
and by choosing the renormalization scale $\mu=\Lambda=m_\kappa$, we 
obtain the following coefficients 
$\alpha_{ij}^{R} = 20/9 h^*_{ki} h_{kj}$ and 
$\alpha_{ij}^{\sigma R}  =  2/3 h^*_{ki} h_{kl}.$ 
 Therefore  the full
amplitude for \mec{} is dominated by the running from the
scale of new physics $\Lambda= m_\kappa$ to relevant scale of the 
process $m_\mu.$  This conclusion is independent on the model as 
long as the effective four-fermion  interactions  exist.
Substituting these results to \eq{tirate} and   
using the present experimental limit for $Ti$
we obtain from \mec{} for $m_\kappa=1$ TeV
\bea
h_{e\mu}h_{ee}^*\;,\;h_{\mu\mu}h_{e\mu}^*\lsim
\frac{6\cdot 10^{-4}}{\sqrt{B}}\,, \;\;\;\;\;\;\;
h_{\tau\mu}h_{e\tau}^*\lsim\frac{9\cdot 10^{-4}}{\sqrt{B}}\,, 
\label{hk++}
\eea
while $\mu\rightarrow e\gamma$ gives
$h_{k\mu}h_{ek}^*\lsim 3\cdot 10^{-3}.$
Here we have introduced a factor $B=R^{present}_{\mu e}/R^{future}_{\mu e}$
which takes into accunt the expected experimental improvements.
The bounds \rfn{hk++} are new limits
on the off-diagonal doubly charged scalar interactions (note that 
tree level $\mu\rightarrow 3e$ probes only $h_{\mu e}h_{ee}^*$). 
The upper bounds \rfn{hk++} 
are going to be improved soon by an order of magnitude, 
$\sqrt{B^{Ti}}\approx 12,$
with the expected \mec{} data.

\section{MSSM without $R$-parity}

Within the MSSM particle content  the gauge invariance and supersymmetry 
allow for the following $R$-violating superpotential \cite{AR}
\beq
W_{R\!\!\!\!/} = \lambda_{ijk} \widehat L_i\widehat L_j\widehat E^c_k +
\lambda '_{ijk} \widehat L_i\widehat Q_j\widehat D^c_k +
\lambda ''_{ijk} \widehat U^c_i\widehat D^c_j\widehat D^c_k
-\mu_i L_i H_2\,,
\label{WRviol}
\eeq
where $\lambda_{ijk} = -\lambda_{jik} $ and 
$\lambda ''_{ijk} = -\lambda ''_{ikj} $.
The $\lambda,\,\lambda '$ and $\mu  $ terms
violate the lepton number, whereas the $\lambda '' $ terms
violate the baryon number by one unit.
The last bilinear term in \Eq{WRviol} gives rise to interesting 
physics which has been studied elsewhere \cite{valle}. 

\begin{table}
\begin{tabular}{|c|c|c|c|}
\hline \hline
& previous bounds & \multicolumn{2}{c|}{\mec\ at loop level$/\sqrt{B}$} \\
\cline{2-4}
& $m_{\tilde f}=$ & \multicolumn{2}{c|}{$m_{\tilde f}=$} \\
&  100 GeV & $\;\;\;\;$ 100 GeV $\;\;\;\;$ & 1 TeV   \\ 
\hline \hline
$|\lambda_{121}\, \lambda_{122}|$ & $6.6\cdot 10^{-7}$ Ref.\cite{CR} &
$4.2\cdot 10^{-6}$ & $3.2\cdot 10^{-4}$ \\
$|\lambda_{131}\, \lambda_{132}|$ & $ 6.6\cdot 10^{-7}$ Ref.\cite{CR} &
$5.3\cdot 10^{-6}$ & $3.9\cdot 10^{-4}$ \\
$|\lambda_{231}\, \lambda_{232}|$ & $ 5.7\cdot 10^{-5}$ Ref.\cite{CH} &
 $5.3\cdot 10^{-6}$ & $3.9\cdot 10^{-4}$ \\
$|\lambda_{231}\, \lambda_{131}|$ & $ 6.6\cdot 10^{-7}$ Ref.\cite{CR} & 
$8.4\cdot 10^{-6}$ & $6.4\cdot 10^{-4}$ \\
$|\lambda_{232}\, \lambda_{132}|$ & $ 1.1\cdot 10^{-4}$ Ref.\cite{CH} &
$8.4\cdot 10^{-6}$ & $6.4\cdot 10^{-4}$ \\
$|\lambda_{233}\, \lambda_{133}|$ & $ 1.1\cdot 10^{-4}$ Ref.\cite{CH} & 
$1.7\cdot 10^{-5}$ & $1.0\cdot 10^{-3}$ \\
\hline \hline
\end{tabular}
\caption{
Upper limits on the products  $|\lambda\lambda| $ testable in \mec\
 for two different 
scalar masses $m_{\tilde f}=$100 GeV and  $m_{\tilde f}=$1 TeV.
The previous bounds scale quadratically with the sfermion mass.
The scaling factor $B$ is defined in the text and currently $B=1\,.$}
\end{table}

\mec\ probes the products of the couplings of the type  
$\lambda\lambda$ only at one loop level,  $\lambda\lambda'$ only at
tree level and  $\lambda'\lambda'$ both at tree and one loop level.
The previous bounds
have been collected and updated in the recent reviews \cite{reviews}.
We have calculated the new loop level bounds in Ref. \cite{HMRS}.
The \mec\ tree level bounds are all taken  from Ref. \cite{kim} assuming 
that there is no cancellations between different contributing terms.
For $\lambda\lambda$-s
comparison with the previously obtained bounds in Table 1 shows that in
three cases out of six our new bounds  are more stringent.
This is because the conversion is enhanced by large logarithms
$\ln (m_f^2/m_{\tilde f}^2).$
If SINDRUM II will reach  $\sqrt{B^{Ti}}=12$ then {\it all}  the \mec\ bounds
will be more stringent than the previous ones.

By far the most stringent constraint on the products
$|\lambda_{122}\lambda'_{211}|,$  $|\lambda_{132}\lambda'_{311}|,$
$|\lambda_{121}\lambda'_{111}|$ and $|\lambda_{231}\lambda'_{311}|$
follows from the tree level \mec\  and is 
$4.0\cdot 10^{-8}$ for  $m_{\tilde f}=100$ GeV.

The bounds on $\lambda'\lambda'$ derived from the tree level
\mec\ are very strong for most of the couplings but not for all of them.
The reason is that for some combinations of the couplings, especially
if the third family squarks are involved, their contribution to the
\mec\ is strongly suppressed by small off-diagonal
CKM matrix elements (as much as $\lambda_W^6,$
where $\lambda_W\sim 0.2$ is the Wolfenstein parameter).
With the present constraints \cite{reviews} only
the bound  $|\lambda'_{223}\lambda'_{133}|\lsim 1.2\cdot 10^{-5}$ 
for the relevant sfermion mass $m_{\tilde f}=100$ GeV
is stronger than the tree level \mec\ bounds.
Importantly, as the result of our calculation \cite{HMRS}
the bounds  $|\lambda'_{232}\lambda'_{132}|\lsim 8.7\cdot 10^{-5}$ and
$|\lambda'_{233}\lambda'_{133}|\lsim 8.7\cdot 10^{-5}$ 
derived from the {\it loop} induced
\mec\  are  more stringent than the ones from the 
tree level \mec.


\section{Conclusions}

In conclusion, using the effective Lagrangian description of new 
physics we show that in  a wide class of models 
loop induced $\mu$--$e$ conversion in nuclei is 
enhanced by large logarithms. With the present upper limits on  
$\mu$--$e$ conversion and $\mu\rightarrow e\gamma$ branching ratios
bounds on new physics (occurring at loop level) derived from these processes
are more restrictive in the case of  $\mu$--$e$ conversion. 
This result is confirmed by explicit calculations in the models with
doubly charged Higgses and MSSM without $R$-parity.
Due to the expected improvements in the sensitivity  of 
already running $\mu$--$e$ conversion experiments this process
will become the most stringent test of muon number conservation.

\end{document}